\documentclass{article}

\usepackage{arxiv}

\usepackage[utf8]{inputenc} 
\usepackage[T1]{fontenc}
\usepackage{graphicx}
\usepackage{tikz}
\usetikzlibrary{arrows,automata}
\usepackage{hyperref}
\usepackage{color}

\newcommand{\tool}[0]{\textsc{pyModRev}}
\title{\tool{}: a Python Tool for Model Revision of Boolean Networks}

\author{
 Pedro T. Monteiro \\
  INESC-ID, Department of Computer Science and Engineering\\
  Instituto Superior Técnico\\
  Universidade de Lisboa\\
  Portugal\\
  \texttt{pedro.tiago.monteiro@tecnico.ulisboa.pt} \\
   \And
 Filipe Gouveia \\
  LASIGE, Department of Informatics\\
  Faculdade de Ciências\\
  Universidade de Lisboa\\
  Portugal\\
  \texttt{jfgouveia@ciencias.ulisboa.pt} \\
}

\begin{document}
\maketitle
\begin{abstract}
Biological regulatory networks can be represented by computational models, which allow the study and analysis of biological behaviours, therefore providing a better understanding of a given biological process.
However, as new information is acquired, biological models may need to be revised in order to also account for this new information.
Current model revision tools are scarce and often lack the flexibility to integrate with broader analysis workflows.
Here, we present \tool{}, an enhanced iteration of the model revision tool \textsc{ModRev}, capable of verifying the consistency of Boolean regulatory models, and finding minimal repairs in case of inconsistency.
\tool{} supports model validation against both steady state observations as well as time-series data, being able to consider different update schemes simultaneously.
\tool{} supports different model formats, and is available as a Python package in PyPI, for easy integration with other model analysis tools, significantly improving accessibility and utility for the logical modelling community.
\end{abstract}


\section{Introduction}

Biological regulatory networks can be represented by computational models, which are of great interest in Systems Biology~\cite{karlebach2008modelling}.
These computational models allow the study and analysis of biological behaviours, therefore providing a better understanding of a given biological process.
However, as new experimental data become available, these computational models may become inconsistent, \emph{i.e.}, they may not be able to reproduce the new information acquired.
In this case, models need to be revised~\cite{gouveia2021phd,gouveia2019revision}.

Repairing an inconsistent model is a hard problem, due to the inherent combinatorial nature associated with all the possible changes that can be made to render a model consistent.
Additionally, model revision tools are scarce and often lack the flexibility to integrate with broader analysis workflows.
In this work, we present \tool{}, an enhanced iteration of the model revision tool \textsc{ModRev}~\cite{gouveia2020modrev}, addressing many of its limitations.



Given a Boolean model and a set of experimental observations, \tool{} determines whether the model explains the observational data.
If inconsistencies are found, it identifies minimal repair operations to fix the model, and generates the corresponding revised model files.
The tool supports model validation (and repair) against both steady state observations and time-series data, considering different update schemes: synchronous, asynchronous, and complete (also known as general asynchronous).
Furthermore, \tool{} is able to validate (and repair) a model considering several simultaneously observational data with different update schemes, which was a limitation of \textsc{ModRev} addressed in this work.
Furthermore, \tool{} can now also consider an observation specifying that a given state cannot be a steady state, repair the model such that the state is no longer steady.

\tool{} is open-source\footnote{\url{https://github.com/ptgm/pymodrev/}} and is openly available as a Python package\footnote{\url{https://pypi.org/project/pymodrev/}}.
Moreover, it supports different model formats, such as \textsc{ModRev}'s ASP-based description~\cite{gouveia2020modrev}, BoolNet format~\cite{muessel2010boolnet}, and GINsim format~\cite{ginsim}, as well as different observation formats: ASP-based description and tabular formats.
Due to its availability as a Python package and its modular implementation, \tool{} can integrate easily with other model analysis tools, significantly improving accessibility and utility for the logical modeling community.

The paper is organised as follows. Section \ref{sec:preliminaries} presents the background and related work. The \tool{} tool is presented in Section \ref{sec:pymodrev}.
Evaluation is described in Section \ref{sec:evaluation}.
We conclude and discuss future work in Section \ref{sec:conclusion}.
Appendix \ref{sec:tutorial} presents a tutorial and a user manual of the tool.
A case study is presented in Appendix \ref{sec:case-study}. 

\section{Preliminaries}
\label{sec:preliminaries}

Biological regulatory networks describe complex biological processes. A regulatory network is composed of a set of biological compounds and corresponding interactions.
Different formalisms are used to build computational models of such networks~\cite{de2002modeling,karlebach2008modelling,siegel2006qualitative,thomas1973boolean}. Here, we consider the Boolean logical formalism~\cite{thomas1973boolean}.

A Boolean logical model is defined by: a set of nodes, representing biological compounds; a set of edges, representing interactions between compounds; and a set of regulatory functions, one for each node (compound).
Each node has a Boolean variable associated with it, representing whether the corresponding compound is present (1, TRUE) or absent (0, FALSE).
Edges have a sign associated, representing positive (activations) or negative (inhibitions) interactions between compounds (nodes).
The value of a node can change over time, depending on the value of its regulators, \emph{i.e.}, nodes with an incoming edge, and the corresponding regulatory function.
In this work, we consider all regulators to be essential and without a dual role, thus regulatory functions are considered to be monotone non-degenerate Boolean functions~\cite{cury2019partial,Cury2025}.

\subsection{Related Work}

In systems biology, when modeling a regulatory network, several works focus on the inference of the computational model given a set of experimental observations and some prior knowledge~\cite{bonesis-CMSB24,Guziolowski2013,Huvar2026,Ostrowski2016,Schaub2014}.
In this work, we focus on the revision process, where we start with a fully instantiated model and a set of observations, and the goal is to assess whether the model can reproduce the given observations, and if not, repair the model.

Current model revision tools are scarce and often lack the flexibility to integrate with broader analysis workflows.
Some approaches to model revision consider different formalisms, such as the Sign Consistency Model formalism, with less expressiveness in regulatory function when compared to the logical formalism~\cite{gebser2010repair,siegel2006qualitative}.
Other approaches have a limitation in the repair operations that can be performed, the definition of the regulatory functions, and the supported dynamic behaviours~\cite{merhej2017repairing}.

\textsc{ModRev}~\cite{gouveia2020modrev} is a model revision tool capable of repairing inconsistent Boolean models, considering steady-state and time-series observations under synchronous and asynchronous update schemes.
When a model is inconsistent with a given set of observations, \textsc{ModRev} finds a minimal set of repair operations to render the model consistent, considering: changing regulatory functions; changing the sign of the edges (change between activations and inhibitions); adding new edges (new interactions between biological compounds); and removing existing edges (removing interactions).
However, \textsc{ModRev} has some limitations that hinder its adoption from the wider community, such has: being developed in C++ with non-easy to install bindings; model and observation formats limited to Answer Set Programming (ASP) direct encodings; an architecture not suitable to accommodate alternative update schemes; among others.


\section{\tool{}}
\label{sec:pymodrev}

\tool{} is an enhanced iteration of the model revision tool \textsc{ModRev}~\cite{gouveia2020modrev}, re-written in Python, minimizing external dependencies, 
and capable of addressing most of the previous tool limitations.
It focus on improving usability, accessibility, easiness of integration with other model analysis tools, and utility for the systems biology community.

Our tool implements the revision method previously used by \textsc{ModRev}~\cite{gouveia2021phd,gouveia2020modrev}.
Given a Boolean logical model and a set of experimental observations, it assesses whether the model is consistent using an ASP-based approach.
In case of inconsistency, it finds minimal sets of repair operations to render the model consistent.
Four repair operations are considered: change a regulatory function; change the sign of an edge (type of interaction); remove an edge (a regulator); and add an edge (a regulator).

\begin{figure}[t]
    \centering
    \begin{tikzpicture}[->,>=stealth', shorten >=1pt, auto, node distance=2cm, semithick,scale=1, every node/.style={transform shape},transform shape]

    \tikzstyle{every state} = [text=black,fill=none,draw]
	
        \draw [fill=white!80!blue] (0.2,2.2) rectangle (1.8,3.8);
		\node at (1,3) {Model};
		
		\draw [fill=white!80!orange] (0.2,0.3) rectangle (1.2,1.7);
		\draw [fill=white!80!orange] (0.5,0.5) rectangle (1.5,1.9);
		\draw [fill=white!80!orange] (0.8,0.7) rectangle (1.8,2.1);

		\node at (1.3,1.6) {Exp.};
		\node at (1.3,1.2) {Obs.};
		
		\draw  [->,thick] (1.8,3) -> (2.6,3);
        \draw  [->,thick] (1.8,1.4) -- (2.3,1.4) -- (2.3,2.6) -> (2.6,2.6);

        \draw [fill=none] (2.6,2.2) rectangle (4.4,3.6);
		\node at (3.5,3.1) {Input};
		\node at (3.5,2.7) {Processing};

        \draw  [->,thick] (3.2,2.2) -> (3.2,1.8);   
        \draw  [->,thick] (3.8,1.8) -> (3.8,2.2);  

        \draw [fill=none] (2.6,1.8) rectangle (4.4,0.5);
		\node at (3.5,1.55) {Transform};
		\node at (3.5,1.15) {(Quine};
        \node at (3.5,0.8) {McCluskey)};
		
        \draw [fill=none,dashed] (2.1,0.3) rectangle (12.4,3.8);
        \node at (11.3,3.5) {(\tool{})}; 
        
        \draw  [->,thick] (4.4,2.5) -> (5,2.5); 
		\draw [fill=none,anchor=center] (5,0.5) rectangle (7,2.8);
		\node at (6,1.5) {Check};
		\node at (6,1) {Consistency};

        \draw  [->,thick] (6,0.5) -> (6,0);
        
        \draw [fill=none] (5,0) rectangle (7,-2); 
        \node at (6,-0.2) {Consistency};
        \node at (6,-0.55) {Output};

        \draw [fill=white!80!green] (5.1,-1) rectangle (6.9,-1.4);
        \node at (6,-1.2) {Consistent};

        \draw [fill=white!60!red] (5.1,-1.5) rectangle (6.9,-1.9);
        \node at (6,-1.7) {Inconsistent};
		
        \draw  [->,thick] (7,1.65) -> (7.5,1.65); 
		\draw [fill=none,anchor=center] (7.5,0.5) rectangle (9.5,2.8);
		\node at (8.5,1.8) {Search};
		\node at (8.5,1.3) {Repairs};

        \draw  [->,thick] (8.5,0.5) -> (8.5,0);
        
        \draw [fill=none] (7.5,0) rectangle (9.5,-2); 
        \node at (8.5,-0.2) {Repair};
        \node at (8.5,-0.55) {Output};
        
        \draw [fill=white!80!yellow] (7.8,-0.8) rectangle (9.2,-1.2);
        \node at (8.5,-1) {Repairs};
        
        \node at (8.5,-1.35) {\ldots};
        
        \draw [fill=white!80!yellow] (7.8,-1.5) rectangle (9.2,-1.9);
        \node at (8.5,-1.7) {Repairs};
		
        \draw  [->,thick] (9.5,1.65) -> (10,1.65);
        \draw [fill=none] (10,0.5) rectangle (12,2.8);
        \node at (11,1.8) {Model};
		\node at (11,1.3) {Repair};

        \draw  [->,thick] (11,0.5) -> (11,0);
        
        \draw [fill=none] (10,0) rectangle (12,-2); 
        \node at (11,-0.2) {Model};
        \node at (11,-0.55) {Output};
        
        \draw [fill=white!80!blue] (10.3,-0.8) rectangle (11.7,-1.2);
        \node at (11,-1) {Model};
        
        \node at (11,-1.35) {\ldots};
        
        \draw [fill=white!80!blue] (10.3,-1.5) rectangle (11.7,-1.9);
        \node at (11,-1.7) {Model};

        \draw [fill=white!90!black] (5,3.1) rectangle (8.5,3.6);
        \node at (6.75,3.35) {\textsc{pyfunctionhood}};
        \draw  [<->,thick] (4.4,3.35) -> (5,3.35);
        \draw  [<->,thick] (9,2.8) -- (9,3.35) -> (8.5,3.35);

        \draw [fill=white!90!black] (5.3,2) rectangle (6.7,2.5);
        \node at (6,2.25) {\textsc{clingo}};

\end{tikzpicture}
    \caption{\tool{} workflow.}
    \label{fig:tool}
    \vspace*{-0.2cm}
\end{figure}

In Figure \ref{fig:tool}, we present an overall workflow of \tool{}, highlighting some of the new features.
In order to improve usability and provide an easier way to integrate with other model analysis tools, \tool{} supports different model formats, supporting not only the original ASP-based format used in \textsc{ModRev} (\texttt{.lp}), but also the BoolNet format (\texttt{.bnet})~\cite{muessel2010boolnet}, and GINsim format (\texttt{.ginml/.zginml})~\cite{ginsim}.
Observational data can be given using the original ASP-based format (\texttt{.lp}), or using well-known tabular formats (\texttt{.csv}, \texttt{.xls}, \texttt{.xlsx}).
For easy integration with other tools or workflows, such as the CoLoMoTo Notebook~\cite{Naldi2015}, different output formats for the information produced by our tool are supported: compact format (easily parsable), JSON format (more verbose but structured), and human-readable. Note that revised models produced as output maintain the format provided in the original model to be revised.

Like in \textsc{ModRev}, regulatory functions are monotone non-degenerated Boolean functions, represented in Canonical Disjunctive Normal Form (CDNF)~\cite{cury2019partial,Cury2025}.
To support different model formats, considering that functions in BoolNet and GINsim are defined as strings representing the Boolean formula, we use the Quine-McCluskey algorithm~\cite{McCluskey1956} to transform into CDNF, before using it in \texttt{pyfunctionhood}. Note that GINsim models relying on parameters for function specification, instead of a function string, are currently not supported.

In \tool{}, the model revision process is now divided into three consecutive phases: consistency check, repair search, and model repair.
The previous tool (\textsc{ModRev}) supported only consistency check and repair search, and without a clear separation between them. 
Here, we introduce the last phase, model repair, to actually apply identified minimal repairs to the original model and generate all the corresponding model files. We also separate each phase to allow stopping the process at any phase, getting the corresponding output information.

The consistency check phase is an ASP-based approach using the ASP solver \texttt{clingo}\footnote{\url{https://pypi.org/project/clingo/}}~\cite{GEBSER2018}.
It produces as output whether the model is consistent, or in case of inconsistency, the information regarding the minimal nodes that need to be repaired, and the experimental observations that originated such inconsistency.
The repair search phase searches for possible repair operations that can render the model consistent, given the possible reasons of inconsistency identified in the previous phase. When considering changing a regulatory function, we want to change it for the closest consistent function considering the family of monotone Boolean functions, since it reduces the changes on the dynamic behaviour~\cite{cury2019partial}. For that, we use \texttt{pyfunctionhood}\footnote{\url{https://pypi.org/project/pyfunctionhood/}}, a library to compute immediate neighbours of Boolean functions~\cite{Cury2025}. This Python library is a corrected version of a previous one (used by \textsc{ModRev}).
This search phase can produce as output the set of minimal repairs found.
The newly added model repair phase produces all the models (in the desired format) that result from applying the minimal repair operations found to render the original model consistent. Note that multiple minimal sets of repairs can be found in the search phase, thus several consistent models can be produced.

In our tool, we support different types of observations: steady-state and time-series. For time-series observations, we support synchronous, asynchronous, and complete (also known as general asynchronous) update schemes.
The observations provided can be fully instantiated observations, in which all state values are defined, or they can have missing values, \emph{i.e.}, observations where some values were not observed.
In case of missing values, our tool assumes they can have any value (0/1), considering all combinations to assess consistency. Note that if an observation has only an initial and final state, with missing values in all the states in between, the tool will consider all possible state transitions, being equivalent to check the reachability between two states.
Enhancing the capabilities of previous work, our tool now supports the definition of not steady states, meaning that the model cannot have them as part of its steady states. If it does, our tool will repair the model so that these states are no longer steady.
Furthermore, our tool can now confront a model with several observational data simultaneously, from different types and update schemes.

In \tool{}, we changed the handling of different update schemes used in \textsc{ModRev} to a more modular implementation. The intention for this effort was to easily associate a possibly different update scheme with each observation, in order to assess the model consistency simultaneously on all observations.
Also, this paves the way to easily extend our tool with future update schemes, such as the Most Permissive update scheme~\cite{bonesis-CMSB24}.



\section{Evaluation}
\label{sec:evaluation}

To assess the correctness of \tool{}, we considered the dataset of corrupted model files created by Gouveia~\cite{gouveia2021phd}. This dataset considers a set of 5 Boolean models, and applies four types of corruptions: change in regulatory function, change in the edge sign, remove a regulator, and add a regulator. For each model, 24 combinations of these corruptions were applied, and for each combination, 100 corrupted instances were generated.
As steady-state observations, the original steady-states of the 5 Boolean models were considered. For dynamical observations, synchronous and asynchronous simulations of 3, 5, 10, 15 and 20 time-steps were also generated.

\tool{} was capable of repairing most models under a time limit of 3600 seconds. Also, all proposed repairs were consistent with the observations and were similar to the repairs proposed by \textsc{ModRev}, showing the correctness of the model revision approach implemented in our tool. 

Additionally, to illustrate the easiness of use of \tool{} to the logical modelling community, in Appendix \ref{sec:case-study} we reproduce the study in \cite{automata24}, where \textsc{ModRev} was used to proposed revisions to the Boolean model on the early differentiation of hematopoietic stem cells, proposed by H\'eraut et al. \cite{Herault2022}, in order to recover desired reachabilities with the asynchronous update scheme.

\section{Conclusion and Future Work}
\label{sec:conclusion}

In this work, we present \tool{}, an enhanced iteration of the model revision tool \textsc{ModRev}~\cite{gouveia2020modrev}, that assesses whether a Boolean logical model of a biological regulatory network is consistent with a set of experimental observations.
We show the correctness of \tool{} with respect to \textsc{ModRev}, and its usability and utility by providing it as Python package in PyPI.
Nevertheless, we intend to improve \tool{} in several aspects.

Currently, for time-series observations, the exact number of time steps must be provided to the tool, making the search incremental.
We intend to relax this constraint to fully support reachability properties without knowing the number of time-steps in between (partial) states.

Also, we already refactored the handling of the update schemes in \tool{} to easily extend to new update schemes.
As future work, we plan to include support for reachability properties using the Most Permissive update scheme~\cite{bonesis-CMSB24}.

We included three different output formats, notably the JSON format, with the aim to include \tool{} within the CoLoMoTo Notebook~\cite{Naldi2015} for easier integration with other analysis pipelines, already used by the community.



\subsubsection*{Funding}
This work was supported by national funds through FCT, Fundação para a Ciência e a Tecnologia, under projects
UID/50021/2025 (DOI: 10.54499/UID/50021/2025),
UID/00408/2025 (DOI: 10.54499/UID/00408/2025),
FCT TENURE program 2023.15441.TENURE.044/CP00003/CT00033,
FCT-Mobility investment 11/C06-i06/2024 (FCT/Mobility/1419033983/2024-25),
and by OSCARS project, funded by the European Commission’s Horizon Europe Research and Innovation Programme under grant agreement No. 101129751.

%
%
%
\bibliographystyle{splncs04}
\bibliography{lib}

\appendix
\section{Tutorial / User manual}
\label{sec:tutorial}

\textsc{pyModRev} is a Python reimplementation of \textsc{ModRev} for automated: i) consistency checking of Boolean logical models against experimental observations, ii) computation of minimal repair operations using Answer Set Programming (ASP) and the \texttt{clingo} solver, and iii) automatic production of the corresponding repaired models.

The tool supports multiple model formats (\texttt{.lp}, \texttt{.bnet}, \texttt{.ginml}/\texttt{.zginml}) for input and output, and multiple observation formats (\texttt{.lp}, \texttt{.csv}, \texttt{.xls}, \texttt{.xlsx}).

\subsection{Preparation}

\subsubsection{Installation}
\label{sec:tutorial-install}

You can install it from source (inside the repository directory), having previously git cloned or downloaded the project:
\begin{verbatim}
pip install .
\end{verbatim}

or install the latest release directly from PyPI:
\begin{verbatim}
pip install pymodrev
\end{verbatim}

\subsubsection{Command line usage}
\label{sec:tutorial-cli}

Using the command line, pass the \texttt{-h} option to list all possible parameters:
\begin{footnotesize}
\begin{verbatim}
pymodrev -h
usage: cli.py [-h] -m MODEL -obs OBS [UPDATER ...] -t {c,r,m}
              [--exhaustive-search] [-s {1,2,3,4}] [-f {c,j,h}]
              [--fixed-nodes FIXED_NODES [FIXED_NODES ...]]
              [--fixed-edges FIXED_EDGES [FIXED_EDGES ...]] [-d]

options:
  -h, --help            show this help message and exit
  -m, --model MODEL     Input model file
  -obs, --observations OBS [UPDATER ...]
                List of observation files and updater pairs.
                Each observation *must* be followed by its updater type. 
                Example: -obs obs1.lp async obs2.lp sync
                Or: -obs obs1.lp async -obs obs2.lp sync
  -t, --task {c,r,m}    Specify the task to perform (default=r):
                            c - check for consistency
                            r - get repairs
                            m - get repaired models
  --exhaustive-search   Force exhaustive search of function
                repair operations (default=false)
  -s, --solutions {1,2,3,4}
                All solutions are optimal w.r.t. number of nodes needing
                repairs. But a solution may be sub-optimal w.r.t. number
                of repair operations.
                    1 - Show only the first ASP optimal solution, which 
                       may be sup-optimal in terms of repairs (fastest)
                    2 - Show only the first ASP&repairs optimal solution
                    3 - Show all optimal solutions (default)
                    4 - Show all optimal solutions, including sub-optimal 
                        repairs (slowest)
  -f, --format {c,j,h}  Specify output format (default=h):
                            c - compact format
                            j - json format
                            h - human-readable format
  --fixed-nodes FIXED_NODES [FIXED_NODES ...]
                        List of nodes ids not to repair.
                        Example: --fixed-nodes A B C
  --fixed-edges FIXED_EDGES [FIXED_EDGES ...]
                        List of edges ids not to repair.
                        Example: --fixed-edges A,B C;D E:F
  -d, --debug           Enable debug mode

\end{verbatim}
\end{footnotesize}

\subsubsection{Observation file formats}
\label{sec:tutorial-observations}

\textsc{pyModRev} supports observations written as ASP facts (\texttt{.lp}) using \textsc{ModRev}'s format~\cite{gouveia2020modrev}, or in a tabular format (\texttt{.csv}/\texttt{.xls}/\texttt{.xlsx}).
Additionally, it steady-state vs. time-series have slightly different formats to account for an explicit mention of time.

\begin{itemize}
  \item \textbf{Steady-state:} first header field empty, first column is the profile identifier.
\begin{verbatim}
,node1,node2,node3
p1,0,1,0
p2,1,1,1
...
\end{verbatim}

  \item \textbf{Time-series:} two empty header fields, where first column is the profile identifier, second column is time.
\begin{verbatim}
,,node1,node2,node3
p1,0,0,1,1
p1,1,1, ,0
p1,2,*,0,0
...
\end{verbatim}
\end{itemize}

Missing values (empty fields, \texttt{*}, \texttt{N/A}, \texttt{NaN}, \texttt{-}) are skipped so that no constraint is generated for that node/time point, and \tool{} can produce repairs instantiated with either value.

\subsection{Command line usage}

\subsubsection{Checking consistency of a model}

To check if a model is consistent with a set of observations, a user can use the following command:

\begin{verbatim}
pymodrev -m examples/boolean_cell_cycle/03/model.bnet \
  -obs examples/boolean_cell_cycle/03/steady.csv steady -t c
This model is inconsistent!
  node(s) needing repair: "p27", "rb", "cdc20", "cycd"
  present in profile(s): "p1"
\end{verbatim}

In this case, it is indicated that the model has four nodes inconsistent in the specific profile, that should be repaired.
If the model is consistent, the tool will report that no repair is needed.

\subsubsection{Computing minimal repair operations}

To compute the minimal repairs on these four nodes, a user can change the \texttt{-t} parameter to ask for a \texttt{r}epair task:
\begin{verbatim}
pymodrev -m examples/boolean_cell_cycle/03/model.bnet \
  -obs examples/boolean_cell_cycle/03/state.csv steady -t r
### Found solution with 4 repair operations.
	Inconsistent node p27.
		Repair #1:
			Change function of p27 to: (!cycb && !cyce) || 
                              (cyca && !cycb && cycd && !p27)
	Inconsistent node rb.
		Repair #1:
			Change function of rb to: (!cycb && cycd && cyce) || 
                             (!cycb && cycd && !p27) || 
                             (!cycb && !cyca)
	Inconsistent node cdc20.
		Repair #1:
			Flip sign of edge (cycb,cdc20) to: positive
	Inconsistent node cycd.
		Repair #1:
			Flip sign of edge (cycd,cycd) to: positive
\end{verbatim}

Notice that to repair this model, one must apply repairs on four different nodes. And on each node there is a single possible repair. In this case, two nodes need to change its function, and the other two only the sign of the regulatory influence.

There are cases where more than one repair per model node is possible, such as the following example using a toy model and asking for sub-optimal repairs.
Here, we can also observe cases where a repair is composed of multiple changes, e.g. the regulatory sign and regulatory function.

\begin{verbatim}
pymodrev -m examples/toy/00/model.bnet \
  -obs examples/toy/00/sync.csv sync -t r -s 4
(Sub-Optimal Solution)
### Found solution with 4 repair operations.
	Inconsistent node v1.
		Repair #1:
			Change function of v1 to: (v3) || (!v2)
			Flip sign of edge (v2,v1) to: negative
		Repair #2:
			Change function of v1 to: (!v3) || (v2)
			Flip sign of edge (v3,v1) to: negative
	Inconsistent node v2.
		Repair #1:
			Change function of v2 to: (v3 && v1)
			Flip sign of edge (v3,v2) to: positive
		Repair #2:
			Change function of v2 to: (!v3 && !v1)
			Flip sign of edge (v1,v2) to: negative
### Found solution with 3 repair operations.
	Inconsistent node v1.
		Repair #1:
			Change function of v1 to: (v3) || (!v2)
			Flip sign of edge (v2,v1) to: negative
	Inconsistent node v2.
		Repair #1:
			Change function of v2 to: (!v3 && v1)
(Sub-Optimal Solution)
### Found solution with 4 repair operations.
	Inconsistent node v1.
		Repair #1:
			Change function of v1 to: (!v3) || (v2)
			Flip sign of edge (v3,v1) to: negative
	Inconsistent node v2.
		Repair #1:
			Change function of v2 to: (v3 && v1)
			Flip sign of edge (v3,v2) to: positive
\end{verbatim}

\subsubsection{Generating repaired models}

A user can manually edit its model using the suggested repairs, or it can use \tool{} to automatically generate the repaired models files. For this, it suffices to change the \texttt{-t} parameter to ask for repaired \texttt{m}odels.
In the optimal solution of the Boolean cell cycle case, since there was only one possible repair per model node, there is a single generated model containing all the repairs.

\begin{verbatim}
pymodrev -m examples/boolean_cell_cycle/03/model.bnet \
  -obs examples/boolean_cell_cycle/03/steady.lp steady -t m
Repaired model: examples/boolean_cell_cycle/03/model_1.bnet
\end{verbatim}

Using the toy example, and asking for sub-optimal repairs, we can see that six repaired models are generated, considering the number of solutions, and the combination of repairs within the model nodes in each solution.

\begin{verbatim}
pymodrev -m examples/toy/00/model.bnet \
  -obs examples/toy/00/sync.csv syncupdater -t m -s 4
Repaired model: examples/toy/00/model_1.bnet
Repaired model: examples/toy/00/model_2.bnet
Repaired model: examples/toy/00/model_3.bnet
Repaired model: examples/toy/00/model_4.bnet
Repaired model: examples/toy/00/model_5.bnet
Repaired model: examples/toy/00/model_6.bnet
\end{verbatim}

\section{Case study}
\label{sec:case-study}

\begin{figure}
    \centering
    \includegraphics[width=0.9\linewidth]{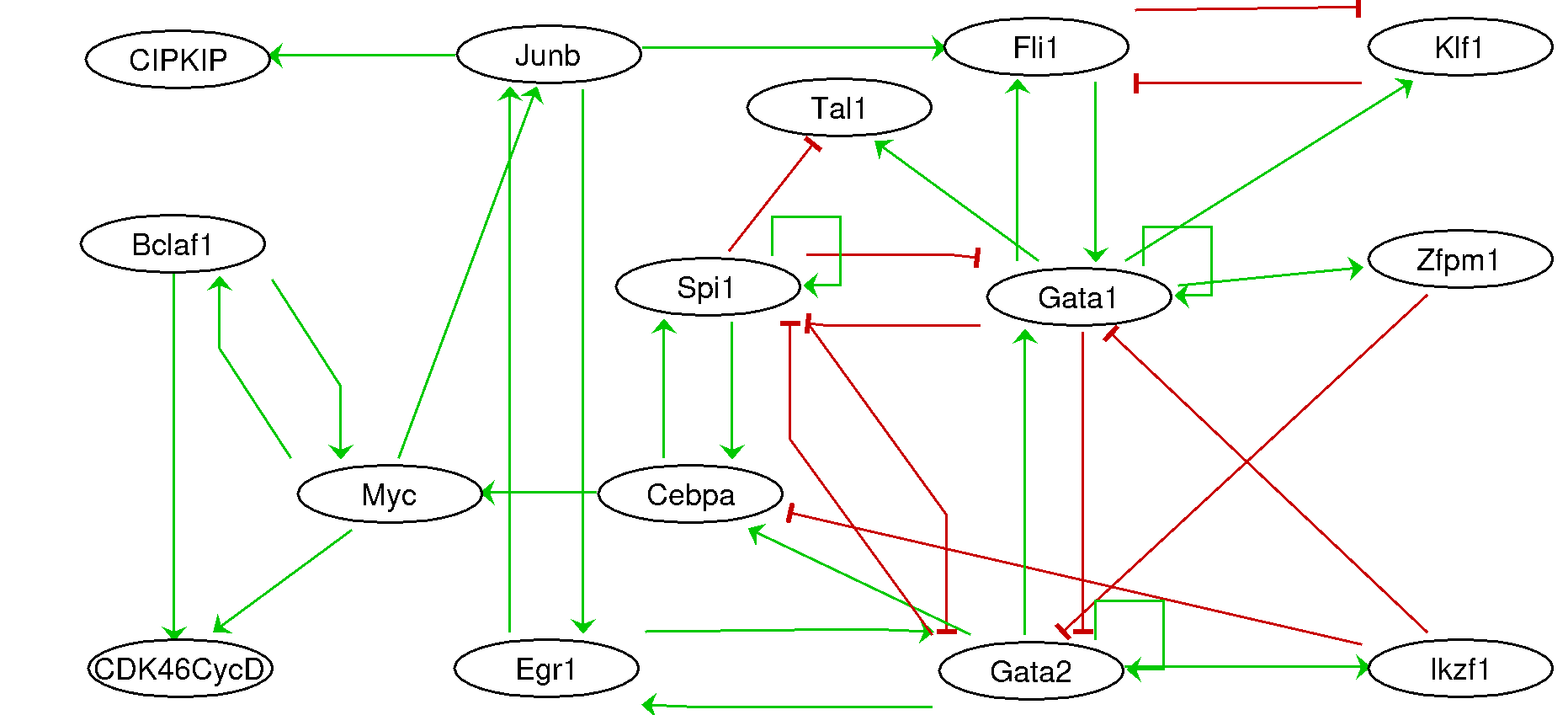}
    \caption{Boolean model of the early differentiation of Hematopoietic Stem Cells (HCS).}
    \label{fig:model}
\end{figure}

H\'eraut et al. \cite{Herault2022} proposed a workflow to build a Boolean model on the early differentiation of hematopoietic stem cells (HSCs), combining gene regulatory network inference methods, single-cell transcriptomic data defining cell (steady) states, and dynamical constraints between cell states.
The resulting model (see Fig.~\ref{fig:model}) was capable of generating 5 cell (steady) states: lymphoid (pLymph), neutrophils and mastocytes (pNeuMast), erythrocytes (pEr), megakaryocytes (pMk), and an inactive state (zeros). It was also capable of generating reachabilities under MP update scheme between some identified cell states and these cell steady states.
Nevertheless, some of these reachabilities were not possible under the asynchronous update scheme.

In \cite{automata24}, the authors utilised \textsc{ModRev} to propose revisions to the original model\footnote{\url{https://ginsim.github.io/models/2023-early-differentiation-hematopoietic-stem-cells/}}, to recover these missing asynchronous reachabilities.
Here, we illustrate how \tool{} can easily be used to recover these asynchronous reachabilities.

We start by defining a tabular \texttt{.xlsx} file with the 5 cell (steady) states:
\begin{verbatim}
,Egr1,Junb,Bclaf1,Myc,Fli1,Gata2,Spi1,Cebpa,Gata1,Klf1,Tal1,\
Ikzf1,Zfpm1,CDK46CycD,CIPKIP
Zero,0,0,0,0,0,0,0,0,0,0,0,0,0,0,0
pEr,0,0,0,0,0,0,0,0,1,1,1,0,1,0,0
pNeuMast,0,0,0,0,0,0,1,1,0,0,0,0,0,0,0
pLymph,0,0,0,0,0,1,1,0,0,0,0,1,0,0,0
pMk,0,0,0,0,1,0,0,0,1,0,1,0,1,0,0
\end{verbatim}

We then check for the consistency of the model against this observation tabular file, which is expected, using the following command with task parameter \texttt{-t c}:
\begin{verbatim}
pymodrev -m hsc.zginml -obs hsc_steadystates.xlsx steady -t c
This model is consistent!
\end{verbatim}

\subsection{Reachability: iHSC to pLymph}

In \cite{automata24} the authors started by verifying the reachability between iHSC and pLymph.
We define this reachability property from state iHSC to state pLymph, with five time steps, in a tabular \texttt{.xlsx} file as follows, where ``iHSC2pLymph'' is the name of the observation profile:
\begin{verbatim}
,,Egr1,Junb,Bclaf1,Myc,Fli1,Gata2,Spi1,Cebpa,Gata1,Klf1,Tal1,\
Ikzf1,Zfpm1,CDK46CycD,CIPKIP
iHSC2pLymph,0,0,0,1,0,1,1,0,0,0,0,1,0,0,0,0
iHSC2pLymph,5,0,0,0,0,0,1,1,0,0,0,0,1,0,0,0
\end{verbatim}

We then check for the consistency of the model against this reachability, and observe that there is a node, in this case ``Spi1'', that needs to be repaired.
\begin{verbatim}
pymodrev -m hsc.zginml -obs hsc_steadystates.xlsx steady \
  -obs hsc_iHSC_2_pLymph.xlsx async -t c
This model is inconsistent!
  node(s) needing repair: "Spi1"
  present in profile(s): "iHSC2pLymph"
\end{verbatim}

To ask for the specific repairs to make the model consistent with both observations, we change the task parameter \texttt{-t r}, and the solutions parameter \texttt{-s 1} (considering sub-optimal repairs - the fastest). One repair with two changes must be applied to node ``Spi1'':
\begin{verbatim}
pymodrev -m hsc.zginml -obs hsc_steadystates.xlsx steady \
  -obs hsc_iHSC_2_pLymph.xlsx async -t r -s 1
### Found solution with 2 repair operations.
  Inconsistent node Spi1.
	Repair #1:
	   Change function of Spi1 to: (!Gata1 && Gata2) || \
                (!Gata1 && Spi1) || (!Gata1 && Cebpa)
	   Flip sign of edge (Gata2,Spi1) to: positive
\end{verbatim}

To generate the repaired model file, we change the task parameter to \texttt{-t m}:
\begin{verbatim}
pymodrev -m hsc.zginml -obs hsc_steadystates.xlsx steady \
  -obs hsc_iHSC_2_pLymph.xlsx async -t m -s 1
Repaired model: hsc_1.zginml
\end{verbatim}

This repaired model file \texttt{hsc\_1.zginml} should be consistent with the observations, which can be confirmed with the following command:
\begin{verbatim}
pymodrev -m hsc_1.zginml -obs hsc_steadystates.xlsx steady \
  -obs hsc_iHSC_2_pLymph.xlsx async -t c
This model is consistent!
\end{verbatim}

\subsection{Reachability: qHSC to pLymph}

Considering this new model, the authors in \cite{automata24} then verified the additional reachability from state qHSC towards pLymph. We define this reachability property in a tabular \texttt{.xlsx} file as follows, where ``qHSC2pLymph'' is the name of the observation profile (note the missing value on node ``zfpm1''):
\begin{verbatim}
,,Egr1,Junb,Bclaf1,Myc,Fli1,Gata2,Spi1,Cebpa,Gata1,Klf1,Tal1,\
Ikzf1,Zfpm1,CDK46CycD,CIPKIP
qHSC2pLymph,0,1,1,0,1,1,1,0,0,0,0,1,0,,1,1
qHSC2pLymph,9,0,0,0,0,0,1,1,0,0,0,0,1,0,0,0
\end{verbatim}

To ask for the specific repairs to make this \texttt{hsc\_1.zginml} model consistent with both the steady states and the ``qHSc2pLymph'' observation, we run the following command with the task parameter \texttt{-t r}, and the solutions parameter \texttt{-s 1} (considering sub-optimal repairs - the fastest). In this case, there is only one minimal repair to be applied to node ``Junb'':
\begin{verbatim}
pymodrev -m hsc_1.zginml -obs hsc_steadystates.xlsx steady \
  -obs hsc_iHSC_2_pLymph.xlsx async
  -obs hsc_qHSC_2_pLymph.xlsx async -t r -s 1
### Found solution with 1 repair operations.
  Inconsistent node Junb.
    Repair #1:
	   Change function of Junb to: (Myc && Egr1)
\end{verbatim}

We generate the repaired model file, changing the task parameter to \texttt{-t m}:
\begin{verbatim}
pymodrev -m hsc_1.zginml -obs hsc_steadystates.xlsx steady \
  -obs hsc_iHSC_2_pLymph.xlsx async
  -obs hsc_qHSC_2_pLymph.xlsx async -t m -s 1
Repaired model: hsc_1_1.zginml
\end{verbatim}

%


\end{document}